\documentclass[useAMS,usenatbib]{mn2e}
\usepackage{graphicx}
\usepackage{graphics}
\usepackage{epsfig}
\usepackage{rotate}

\def\ar{{$\rightarrow$}}

\title[On the 3D dynamics and morphology of inner rings]{On the 3D dynamics 
and morphology of inner rings}
\author[Patsis et al.]{P.A. Patsis,$^{1}$\thanks{E-mail:
ppatsis@cc.uoa.gr (PAP); hskokos@cc.uoa.gr (ChS);
lia@oamp.fr(EA)}\thanks{Visiting Astronomer, Observatoire de
Marseille} Ch. Skokos$^{1}$\footnotemark[1]\footnotemark[2] and E.
Athanassoula$^{2}$\footnotemark[1]\\
$^{1}$Research Center for Astronomy, Academy of Athens, Anagnostopoulou 14, GR-
10673, Athens, Greece\\
$^{2}$Observatoire de Marseille, 2 Place Leverrier, F-13248,
Marseille, Cedex 4, France}

\begin{document}

\date{Accepted ???. Received ???; in original form ???}

\pagerange{\pageref{firstpage}--\pageref{lastpage}} \pubyear{2003}

\maketitle

\label{firstpage}

\begin{abstract}
  We argue that inner rings in barred spiral galaxies are associated with
  specific 2D and 3D families of periodic orbits located just beyond the end
  of the bar. These are families located between the inner radial
  ultraharmonic 4:1 resonance and corotation. They are found in the
  upper part of a type-2 gap of the x1 characteristic, and can account for the
  observed ring morphologies without any help from families of the x1-tree.
  Due to the evolution of the stability of all these families, the ring shapes
  that are favored are mainly ovals, as well as polygons with `corners' on the
  minor axis, on the sides of the bar. On the other hand pentagonal rings, or
  rings of the NGC~7020 type hexagon, should be less probable. The orbits that
  make the rings belong in their vast majority to 3D families of periodic
  orbits and orbits trapped around them.
\end{abstract}


\section{Introduction}

Rings are often observed in barred galaxies and can be spectacular structures.
They come in three kinds: nuclear rings, near the nucleus of the galaxy, inner
rings, surrounding the bar, and outer rings, of a considerably larger
diameter. In this paper we focus our attention to the study of the {\em inner}
rings.  \citet[hereafter B95]{bu95} has made a statistical study of 3692 ring
galaxies in the southern hemisphere, as well as more in depth studies of
specific objects (e.g. \citealt*{buea01, buea99, bupu98,buea96}). He found
that inner rings are a frequently encountered feature of barred galaxies.
Their mean axial ratio is $0.84 \pm 0.10$, elongated along the bar major axis.
This ratio, however, varies from one galactic type to another as well as among
different ring morphologies \citep[see Table VII in][]{buco96}. In some cases
of typical inner rings the axis ratio can be as low as 0.63 (NGC~6782) or even
0.49 (A0106.7-3733) \citep{bu86}. B95 mentions a dozen galaxies for which
there could be an intrinsic misalignment between the bar and the inner ring.
Because of projection effects, however, spectroscopic measurements are
necessary in order to confirm this misalignment. In the vast majority of cases
the inner ring encircles the bar and touches its extremities.  There are,
however, cases where the bar underfills what appears to be an inner ring.
\citet{bu86} mentions 30 such cases in his sample of 1200 objects. Two typical
examples are also given by B95, namely NGC 7098 and NGC 3450.

The link between rings and resonances has been made already in the
'70s \citep[e.g.][]{ss76}. It took, however, the advent of test
particle hydrodynamic simulations in the '80s to establish it
fully. In particular, \citet{sch81, sch84a} followed the response
of gas, modeled by sticky particles, to a bar forcing. He found
that ring structures can form, and that, as in real galaxies, they
can be classified into outer, inner and nuclear. Schwarz linked
the outer rings to the outer Lindblad resonance (OLR), the inner
ones to the inner 4:1 resonance or inner ultraharmonic resonance
(iUHR), and the nuclear ones to the inner Lindblad resonance
(ILR). The first statistical arguments came from \citet{athea82}
who, using the sample of \citet{dvb}, showed that the ratios of
inner to outer ring sizes is compatible with the outer ring being
at the OLR and the inner one at corotation, or at the iUHR.  This
statistics has been since repeated (B95), using much better
observational samples, but the result did not change.

Rings form by gas accumulation at resonances \citep[e.g.][and
references therein]{bu99}, and this explains their predominance on
images of galaxies which show best the sites of population I
objects. It is, however, now clear \citep{ath92b} that the gas
response is determined to a large degree by the periodic orbits in
the underlying gravitational potential. For this reason,
understanding the orbital dynamics of rings is essential not only
for understanding their morphology {\em per se}, but also for
understanding properties related to the gas component, e.g. star
formation.

One of the main uses of orbital structure studies is that they provide
information on the orbits that are the backbone of the various galactic
structures, and thus on their morphologies.  Studying the orbital structure in
an appropriate Hamiltonian system we get the periodic orbits that could be
responsible for the appearance of morphological features of real galaxies. We
are thus interested only in stable periodic orbits, since they trap around
them the regular orbits of the system. While in 2D bar models the backbone of
the bar is a single family, namely the family x1 \citep{gco83}, in 3D models
the orbits supporting the bar are related to a whole tree of families of
periodic orbits, which we called `x1-tree' \citep*{spa02a}.  Under the
assumption that bars are fast, their end should be close and within corotation
\citep{gcoxx}. Due to the fact that inner rings surround the bar, one
intuitively links them to stable orbits not belonging to the x1-tree, but
occupying the area just beyond the bar in the real space.

Although an association of {\em a} resonance location to an inner
ring can be easily done in a first approximation, there is so far
no detailed study relating the various observed morphologies of
inner rings to the orbital dynamics at their region. As Buta
remarks ``...rings include information on ... the properties of
periodic orbits''\citep{bu02}. This information has not been
registered until now, especially for 3D barred models, and this is
exactly what gave us the motivation for the present paper.

Inner rings can be either smooth, in which case we refer to them with the
generic name of `oval', or they can have characteristic `angles', or `kinks',
or `corners' in their outline. These can be more or less strong and, depending
on their strength, may give the ring a polygonal, rather than oval, geometry.
They occur at characteristic locations along the ring. Thus such `corners', or
`density enhancements' often occur near the bar major axis, giving the ring a
`lemon-like' outline. In other cases, we observe that the ring is squeezed
along the major axis of the bar, so that the tips of the `lemon' structure
vanish, the ring becomes rather polygonal-like with sides roughly parallel to
the bar's minor axis at its apocentra, and `corners' close to the minor axis.
Typical examples are UGC~12646 \citep[e.g][-- Fig.~17]{buco96}, IC~4290
\citep[see the distribution of the HII regions in][-- Fig.~8]{buea98},
NGC~3351 \citep[e.g.][ panels 168, 170]{sabe94}.  In such cases it is
reasonable to speak roughly about a hexagon with two sides parallel to the
minor axis of the bar.  The above mentioned objects (UGC~12646, IC~4290 and
NGC~3351) are typical examples. Nevertheless, in most cases all these features
characterize only parts of the rings and perfect symmetry, e.g.  with respect
to the bar major axis, is not always observed.

Despite the complexity of the observed structures of the inner rings, and the
fact that they are located outside the bar, they have been in general vaguely
associated with orbits at the inner 4:1 resonance \citep{sch84, bu99, bu02}.
\citet{sch84} had already realized the presence of squeezed ovals with
corners. He was, however, able to reproduce them only in the case of weak
bars, or by invoking a lens-like component \citep{sch85}. \citet[--
Fig.3]{bu02} invokes combinations of diamond- with barrel-like orbits to
explain ring morphology. In particular barrel-like orbits have been considered
necessary in order to provide cloud--cloud collisions on the sides of the bars
\citep[the regions where the cloud--cloud collisions happen are indicated with
``C'' in][]{sch84}. These orbits help in the formation of rings which have a
hexagonal rather than diamond-like geometry and have two of the sides more or
less parallel to the minor axis of the bar.

Note that neither every diamond-like, nor all hexagonal orbits are necessarily
related with inner rings. E.g. the planar diamond-like x1 orbits in Fig.~62 in
\citet{buco96}, taken from \citet{gcog89}, are far away from the end of the
bar, which is close to corotation. We also note that hexagonal rings with
cusps on the major axis of the bar, and thus with two sides parallel to the
major axis of the bar, are rare. There is only a notable example of this
morphology, namely NGC~7020, studied by \citet{bu90}.

In the present paper we study the orbital behavior at the region
where the appearance of inner rings is favored. We do this in the
case of 3D Ferrers bars. Studying the energy width over which the
various families of periodic orbits exist, their morphology, and
mainly their stability, we identify the orbital behavior in four
archetypical morphologies of inner rings. In section~\ref{Morph}
we present these structures in the cases of four galaxies. In
section~\ref{mode} we give a brief description of the model we use
for our orbital calculations, while the families which build the
rings are described in section~\ref{orb}.  Finally we discuss our
results in section~\ref{blabla} and enumerate our conclusions in
section~\ref{conc}.

\section{Rings morphology}\label{Morph}
\begin{figure}
\centerline{\includegraphics[scale=0.4] {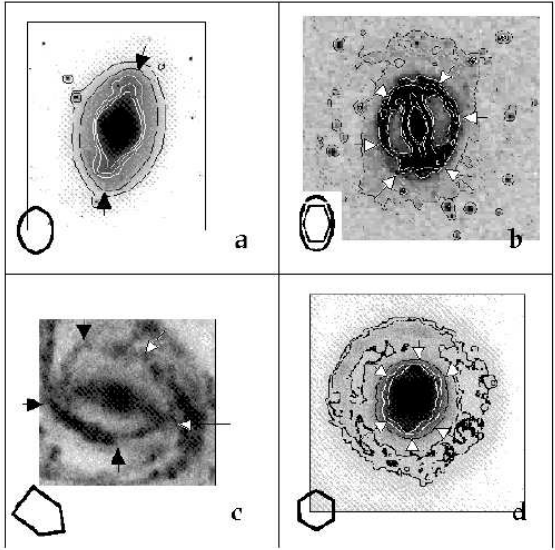}}
\caption{Examples of the four ring morphologies discussed in the
paper. (a) The oval with
  cusps on the major axis (NGC~6782), (b) a
  typical oval-polygonal ring with sides parallel to the minor axis of the bar
  close to the ends of the bar
  (IC~4290), (c) a pentagonal ring in NGC~3367 and (d) the hexagonal
  ring with cusps on the major axis
  (NGC~7020). At the bottom left corner of each panel there is a sketch
  indicating schematically the corresponding morphology.}
  \label{rgal}
\end{figure}

Let us first introduce four inner ring morphologies. Two of them
are typical ovals and the two other are exceptional cases of inner
rings. Although the images\footnote{The image of NGC~6782
originates from an ScTScI Hubble Space Telescope image. It can be
viewed in high resolution at several web sites. The three others
are DSS images. The image of NGC~7020 in Fig.~\ref{rgal}d has been
deprojected. The deprojected, as well as other images of the
NGC~7020 hexagon can be seen in \citet{bu90}} shown in
Fig.~\ref{rgal} are not of high resolution, they are nevertheless
able to demonstrate the features we refer to. Most inner rings are
oval, sometimes with a somewhat lemon shape because of density
enhancements at the bar major axis. We illustrate this latter type
in Fig.~\ref{rgal}a with an image of NGC~6782. The isophotes peak
on the major axis, indicated with arrows on the second outermost
isophote, giving to the ring a cuspy or pointed oval morphology.
There are no ``corners'' close to the minor axis of the bar, which
would have given to the ring a diamond shape.  In Fig.~\ref{rgal}b
we show IC~4290 which has an oval ring with characteristic breaks
or corners. These are indicated with arrows on the overplotted
isophotes. The outermost isophote is again oval-shaped and even
slightly pointed close to the major axis.  The oval, however, is
not smooth and forms, or gives the impression of, angles. One can
also discern a structure in the form of an \mbox{``{\sf
    T}\hspace{-2.5mm}$\perp$''} with the vertical part of the symbol
being the bar.  The two other rings given in Fig.~\ref{rgal} are
exceptional cases.  The ring in Fig.~\ref{rgal}c has an almost
pentagonal geometry. It is observed in an image of NGC~3367. It
can be described as a pseudo-ring, since the ring appears to be
made of segments of spiral arms. Finally in Fig.~\ref{rgal}d we
give a deprojected image of NGC~7020 \citep{bu90}, which has a
type of hexagonal ring with cusps on the major axis of the bar and
two sides parallel to it. The ``corners'' are indicated with
arrows. Of course the geometrical shapes we use above are
idealized and represent the complicated morphology of inner rings
{\em only schematically}. They are, nevertheless, most useful as
guiding lines, in the same way as perfect ellipsoids are useful
for describing bars.

\section{Model}\label{mode}
To calculate the 3D ring orbits we use the fiducial case of the
model described in detail in \citet{spa02a}. It consists of a
Miyamoto disc, a Plummer bulge and a Ferrers bar. The potential of
the Miyamoto disc \citep{mina75} is given by the formula:

\begin{equation}
\label{potd} \Phi
_{D}=-\frac{GM_{D}}{\sqrt{x^{2}+y^{2}+(A+\sqrt{B^{2}+z^{2}})^{2}}}.
\end{equation}
\( M_{D} \) represents the total mass of the disc, $G$ is
the gravitational constant, and $A$ and $B$ are scalelengths such that the
ratio $B/A$ gives a measure of the 
flatness of the model.

The bulge is a Plummer sphere, i.e. its potential is given by:
\begin{equation}
\label{pots} \Phi
_{S}=-\frac{GM_{S}}{\sqrt{x^{2}+y^{2}+z^{2}+\epsilon _{s}^{2}}},
\end{equation}
where \( \epsilon _{s} \) is the bulge scale length and \( M_{S}
\) is its total mass.

Finally, the bar is a triaxial Ferrers bar with density \( \rho
(x) \):
\begin{equation}
\label{densd} \rho (m)=\left\{ \begin{array}{lcc}
\displaystyle{\frac{105M_{B}}{32\pi abc}(1-m^{2})^{2}} & {\mbox
for} &
m \leq 1\\
 & & \\
\displaystyle{0} & {\mbox for}  & m>1
\end{array}\right. ,
\end{equation}
where
\begin{equation}
\label{semiaxis}
m^{2}=\frac{y^{2}}{a^{2}}+\frac{x^{2}}{b^{2}}+\frac{z^{2}}{c^{2}}\,
\, ,\, \,
 \,
 a>b>c.
\end{equation}
In the above \( a \), \( b \), \( c \) are the principal
semi-axes, and \( M_{B} \) is the mass of the bar component. The
corresponding potential \( \Phi _{B} \) and the forces are given
in \citet{pf84} in a closed form, well suited for numerical
treatment. For the Miyamoto disc we use A=3 and B=1, and for the
axes of the Ferrers bar we set a:b:c = 6:1.5:0.6. The masses of
the three components satisfy \( G(M_{D}+M_{S}+M_{B})=1 \).  We
have $GM_{D}=0.82$, $GM_{S}=0.08$, $GM_{B}=0.10$ and $\epsilon
_{s}=0.4$.

The length unit is taken as 1~kpc, the time unit as 1~Myr and the
mass unit as $ 2\times 10^{11} M_{\odot}$. The bar rotates with a
pattern speed $\Omega_{b}$=0.054 around the $z$-axis, which
corresponds to 54~km~sec$^{-1}$~kpc$^{-1}$, and places corotation
at 6.13~kpc.

\section{Orbits}\label{orb}
The Hamiltonian governing the motion of a test-particle in our
system can be written in the form:
\begin{equation}
H= \frac{1}{2}(p_{x}^{2} + p_{y}^{2} + p_{z}^{2}) +
    V(x,y,z) - \Omega_{b}(x p_{y} - y p_{x})    ,
\end{equation}
where $p_{x},~ p_{y},$ and $p_{z}$ are the canonically conjugate
momenta of $x$, $y$ and $z$ respectively and $V(x,y,z)$ is the
total potential of the combined three components of the model:
disc, bar and bulge. We will hereafter denote the numerical value
of the Hamiltonian by $E_j$ and refer to it as the Jacobi constant
or, more loosely, as the `energy'.

A periodic orbit in a 3D Hamiltonian system can be either stable
($S$) or exhibit various types of instability. It can be simple
unstable ($U$), double unstable ($D$) or complex unstable
($\Delta$) \citep[for definitions see][chapter 2.11]{gcobook}.
The method used for finding periodic orbits and their stability is
described in detail in \citet{spa02a} and in references therein.
Periodic orbits are grouped in families, along which the energy
$E_j$ changes and the stability of the orbits may also change. The
S $\rightarrow$ U transitions of periodic orbits are of special
importance for the dynamics of a system, since in this case a new
stable family is generated by bifurcation.  A useful diagram for
presenting the various periodic orbits of the system is the
`characteristic' diagram \citep[section 2.4.3]{gcobook}. It gives
the $x$ coordinate of the initial conditions of the periodic
orbits of a family as a function of their Jacobi constant $E_j$.
In the case of orbits lying on the equatorial plane ($z=p_z=0$)
and starting perpendicular to the $x$ axis ($p_x =0$), with $y=0$
and $p_y > 0$, we need only one initial condition, $x$, in order
to specify a periodic orbit on the characteristic diagram. Thus,
the initial conditions of such orbits are fully determined by a
point on the characteristic diagram.  Initial conditions of orbits
which stay on the equatorial plane but do not start perpendicular
to the $x$ axis and of orbits which do not stay on the equatorial
plane are not fully defined by their $x$ coordinate. In these,
more general cases, the representation on the $(E_j,x)$ diagram is
not very useful.

In our model, the highest energy orbits that support the bar can
be found at the region of the radial 4:1 resonance \citep{spa02a,
psa03}. Since the rings surround the bars, we first look for
families of periodic orbits beyond the energy which corresponds to
this resonance and we follow their stability as we approach
corotation.  \citet{ath92a} has found several families of orbits
beyond the 4:1 resonance in 2D Ferrers bars (see Figs.~2 and 3 of
that paper). We did not know, however, which of those could trap
matter around them and support ring structures, because their
stability was not studied in that paper. We address this
problem here in the more general case of a 3D Ferrers bar.

\subsubsection{The f-group}
In all 3D Ferrers bars studied by \citet*{spa02a,spa02b} the characteristic of
the x1 family has $x$ values which increase with $E_j$ until a maximum is
reached, at which point the $x$ values start decreasing. Beyond the local
maximum we find the characteristics of other families, which for larger $E_j$
have initial $x$'s which are larger than those of x1. In other words, we have
type-2 gaps \citep{gcog89}.  In 2D models of barred galaxies the family found
beyond the local maximum is the family x$_1$(2) \citep[Fig.~3.8b in]{gcobook}.
In our 3D models the first family we find beyond the gap is a 2D 4:1 family,
which we call f. Its morphology is rhomboidal as the morphologies of the
corresponding families of the 2D models in \citet{gcog89} and \citet{ath92a}.
However, in the 3D model, it is accompanied by a forest of bifurcating
families. We call this group of families of periodic orbits the `f-group'.

\begin{figure}
\centerline{\includegraphics[scale=0.4] {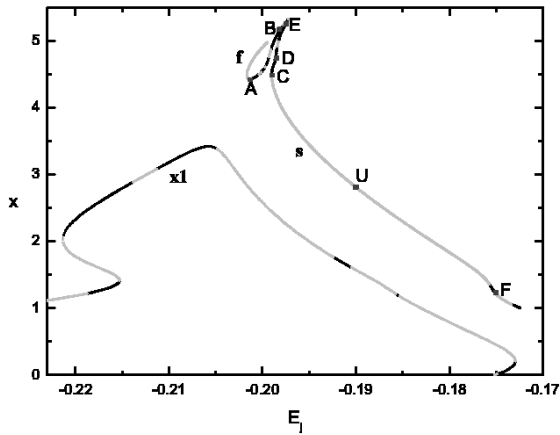}} \vspace{0cm}
\caption[]{Characteristic diagram of families x1, f and s. We
observe a
  typical type-2 gap at the radial 4:1 resonance region.
 Stable parts are indicated with black lines and unstable ones with grey
 lines. The capital letters A to F, as well as U, indicate the location of
 specific orbits
 which are discussed further in the text. At this diagram L$_{4,5}$ would be
  at $\approx$ ($-0.1955, 6.0$).
 }
\label{f-c}\end{figure}

In Fig.~\ref{f-c} we give the ($E_j, x$) characteristic of f, which is the
small curve above the x1 characteristic indicated with f.  The lower part of
the curve, i.e. smaller $x$ for the same $E_j$, is composed mainly of stable f
orbits. Nevertheless, instability strips do exist, but are not important. The
upper part of the curve is mainly composed of unstable orbits. In the figure
black segments indicate stability and grey instability. For the sake of
clarity we do not include in the figure the projections of the characteristics
of the 2D and 3D families bifurcating from f. 

\begin{figure}
\centerline{\includegraphics[scale=0.4] {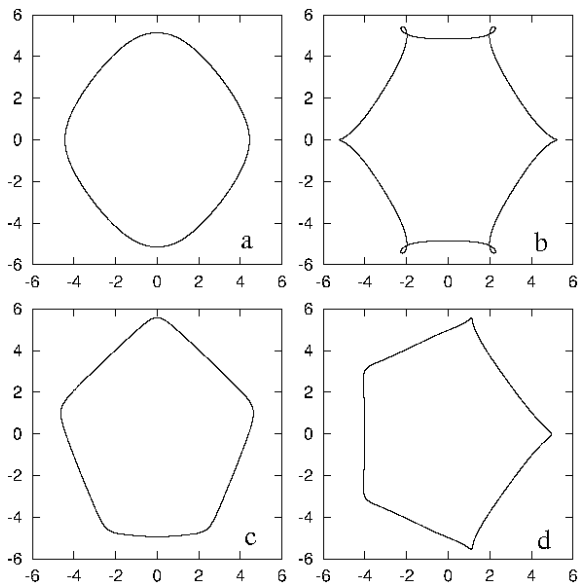}}
\caption[]{ 2D f-orbits. In (a) and (b) we have
stable f orbits located at points ``A'' and ``B'' on the f
characteristic in Fig.~\ref{f-c}. In panels (c) and (d) we give
typical stable fr1 and fr2 orbits. In this and subsequent figures
the bar major axis lies along the $y$-axis and the units are in
kpc. } \label{of2d}
\end{figure}

\begin{figure}
\centerline{ \rotate[r]{\includegraphics[scale=0.4] {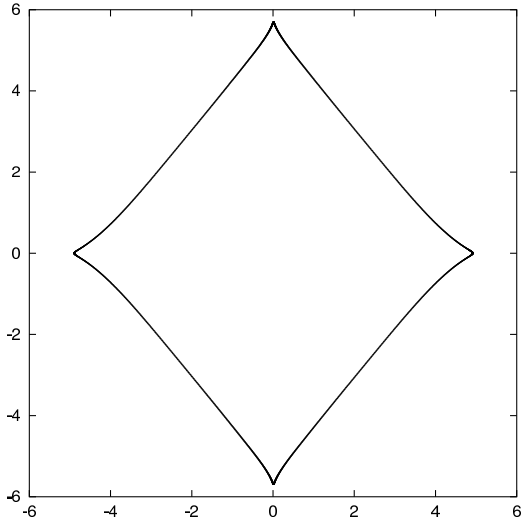}}}
\vspace{-0.2cm} \caption[]{A 2D unstable f-orbit from the upper part
of the f-characteristic
  (drawn with a grey line in Fig.~\ref{f-c}).
 }
\label{urho}
\end{figure}

The `mother' planar 2D f-family starts existing at about $E_j \approx
-0.20166$, and the morphology of its orbits is as in Fig.~\ref{of2d}a. It
remains basically stable until $E_j \approx -0.198$, with only small
instability regions in between. With increasing energy, the morphology of the
f orbits changes and becomes hexagonal (Fig.~\ref{of2d}b). We thus have a
smooth transition from a basically 4:1 to a 6:1 morphology. The two points
marked ``A'' and ``B'' on the stable part of the characteristic indicate the
location of the two f-orbits shown in Fig.~\ref{of2d}a and Fig.~\ref{of2d}b,
respectively. Beyond $E_j \approx -0.198$ the orbital multiplicity, i.e. the
number of intersections of the periodic orbit with the $x$-axis before it
closes, changes.  At an S\ar U\ar S transition close to $E_j \approx -0.2004$
we have two 5:1-type 2D families bifurcated from f. Namely, fr1
(Fig.~\ref{of2d}c) at an S\ar U transition, and fr2 (Fig.~\ref{of2d}d) at the
nearby U\ar S transition. We note that fr1 orbits do not start perpendicular
to the $x$ axis. For these and all the rest of the bifurcating families in the
paper we follow the nomenclature rules introduced in \citet{spa02a}. Soon
after its bifurcation, however, the stable fr1 family becomes unstable, while
the initially unstable fr2 becomes stable. The f orbits along the unstable
part of the characteristic are diamonds with cusps on the axes
(Fig.~\ref{urho}). These orbits are characterized by segments, which resemble
straight line sides.


\begin{figure}
\centerline{\rotate[l]{\includegraphics[scale=0.9] {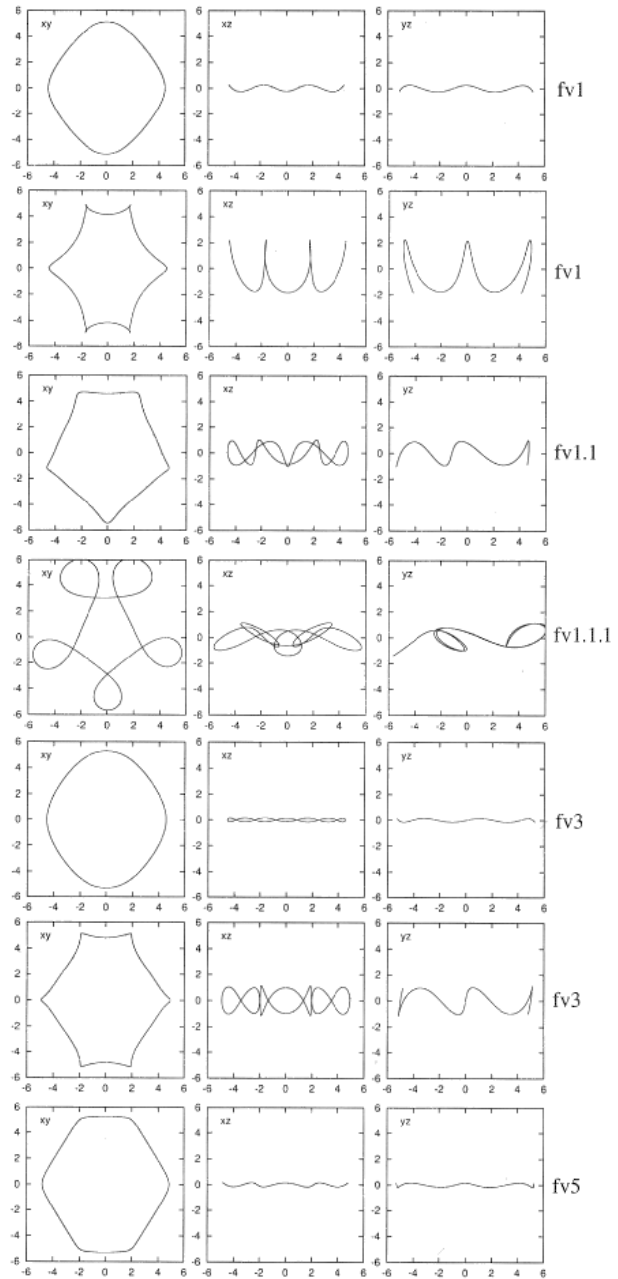}}}\vspace{-0.4cm}
\caption[]{Stable 3D f-orbits. The names of the families are noted
on the right
  side of each set of projections. The projection planes are indicated in the
  upper left corner of each frame.}
\label{of3d}
\end{figure}

Stable 3D families are introduced at S\ar U transitions of family
f at $E_j \approx -0.2016$, $-0.2002$ and $-0.1993$, and are the
families fv1, fv3 and fv5, respectively. These families, together
with their stable bifurcations, contribute significant stable
parts to the area. The three projections of five orbits of 3D
families of the f-group are given in Fig.~\ref{of3d}. The two
orbits at the top of the figure belong to family fv1, for $E_j=
-0.201$ and $-0.181$ respectively, and show that the morphological
evolution of the $(x,y)$ projection of this family follows the
morphological evolution of the 2D family f. The third and fourth
rows from the top give the morphology of orbits from two
bifurcations of family fv1, which in turn has bifurcated from
family f. They are introduced in the system as stable, but do not
have large stability parts, since they become soon unstable.  The
fifth and the sixth orbits from the top belong to family fv3, and
are given for $E_j= -0.200$ and $-0.193$, respectively. The
morphology of their $(x,y)$ projections is again similar to the
morphology of a 2D f-type family (Fig.~\ref{of2d}a,b) of the same
energy. Finally the last orbit is an example of an orbit from
family fv5, close to its bifurcating point. We note that fv5 becomes complex
unstable at $E_j = -0.19697$. At this energy it has already large
deviations above and below the equatorial plane, i.e. large $\overline{|z|}$
values, and thus it would not contribute significantly to the local disc
density, even if it was stable.

\subsubsection{The s-group}
\begin{figure*}
\centerline{\includegraphics[scale=0.8] {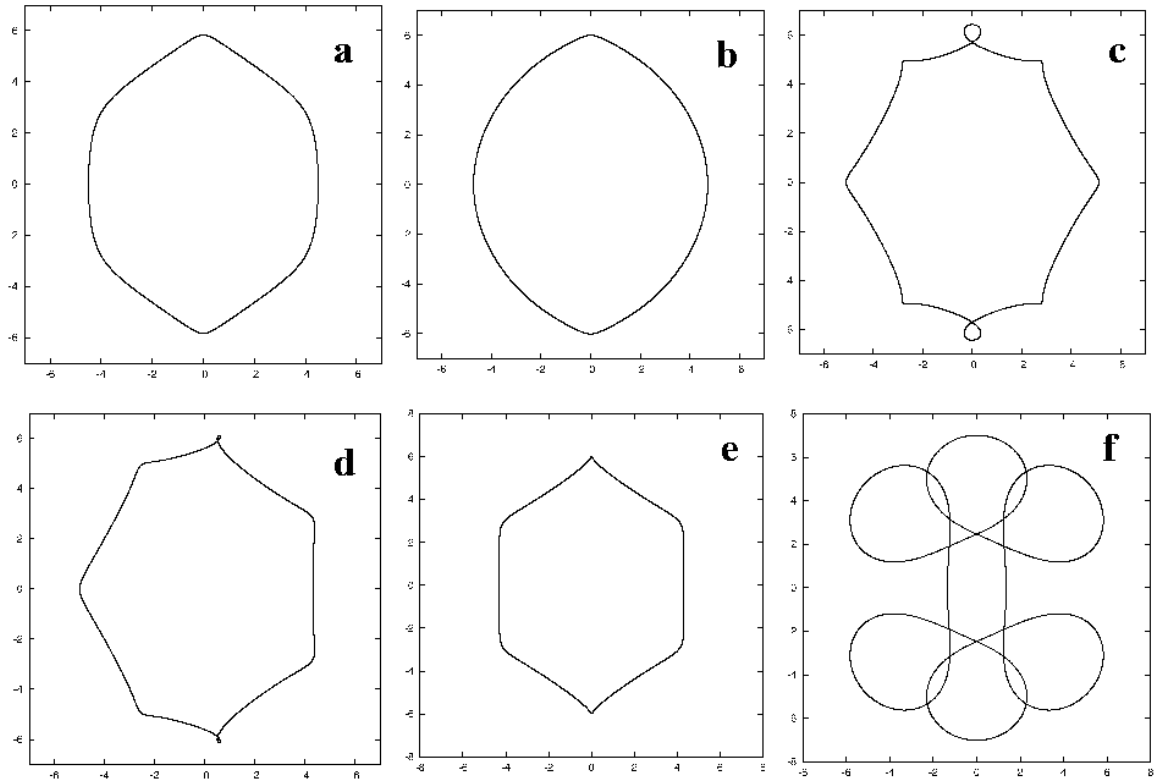}}
\vspace{-0.2cm}
\caption[]{ 2D
orbits of the s-group. From (a) to (b) and then to (c) we see the
evolution of the orbital morphology along the stable branch of the
s characteristic as we move from ``C'' to ``D'' to ``E''
(Fig.~\ref{f-c}). In (d) we observe a planar stable sr1 orbit, in
(e) an unstable s orbit (point U in Fig.~\ref{f-c}), and in (f) a
stable s orbit with loops (point F in Fig.~\ref{f-c}).  }
\label{s2d}
\end{figure*}

In Fig.~\ref{f-c} we see that, below and to the right of the
characteristic of f, there is another, longer curve, roughly
speaking parallel to the decreasing branch of x1. This is the
characteristic of a 2D 6:1 family, which we call ``s''. It exists
for $E_j > -0.1990$. As in the case of the f-group, we have here a
group of families of periodic orbits (s-group). The group includes
s and several 2D and 3D families bifurcated from it. The
stable part is the one with the larger $x$'s for the
same energy. As in the previous case, along the stable part we
have a morphological evolution of the orbits and tiny instability
strips, where new families are born. We can follow the
morphological evolution of the s family of orbits in
Fig.~\ref{s2d} from (a) to (b) and to (c). They correspond to the
points labeled on the characteristic of s in Fig.~\ref{f-c} with
the capital letters ``C'', ``D'' and ``E'', respectively. Along the
characteristic of s we have a smooth transition from a basically 6:1 to an 8:1
morphology.  An S\ar U transition at $E_j \approx -0.19876$
introduces in the system the planar stable family sr1
(Fig.~\ref{s2d}d), which is a 7:1 type orbit. Along the long
unstable part of the s characteristic, orbits are hexagonal with
cusps on the major axis of the bar (Fig.~\ref{s2d}e). Family s has
another stable part at the largest $E_j$'s and small $x$ values
(see Fig.~\ref{f-c}). However, the morphology of the s orbits at
these energies is characterized by big loops (Fig.~\ref{s2d}f), so
that they cannot be used to build rings.

\begin{figure}
\centerline{\rotate[l]{\includegraphics[scale=0.6] {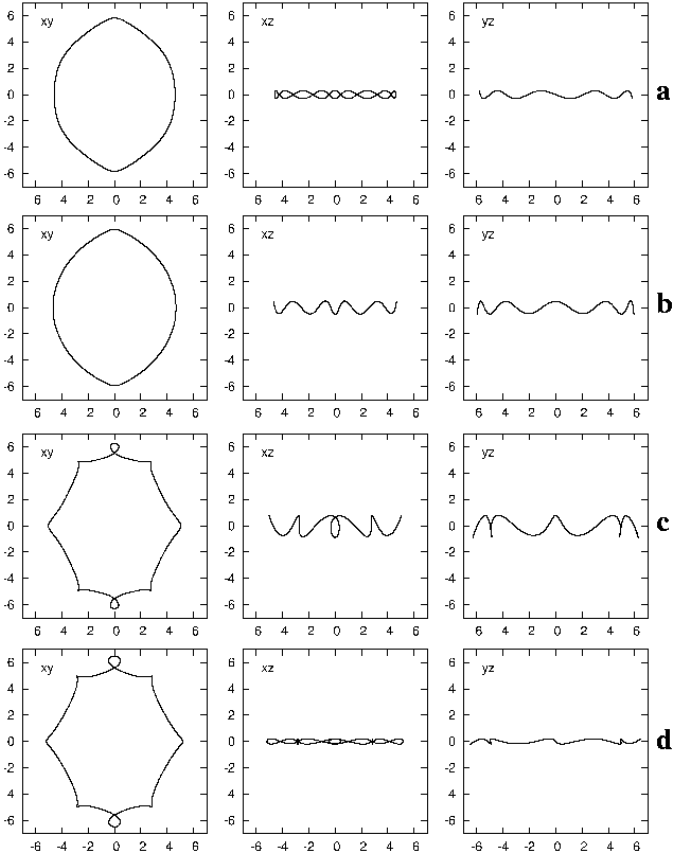}}}
\caption[]{Stable 3D orbits of the s-group. (a) orbit from the sv1
family, (b) sv3 orbit close to the bifurcating point, (c) sv3
orbit away of the bifurcating point, (d) orbit from the sv5
family.} \label{s3d}
\end{figure}

At three other S\ar U transitions with tiny unstable parts, at
$E_j \approx -0.19892, E_j \approx -0.19869$ and $E_j \approx
-0.1984$ respectively, three stable 3D families bifurcate. From
smaller to larger energies, we call them sv1, sv3 and sv5.
Fig.~\ref{s3d} gives the morphology of these families. In
Fig.~\ref{s3d}a we observe the three projections of sv1 close to
its bifurcating point. In Fig.~\ref{s3d}b we give sv3 also close
to its bifurcating point and in Fig.~\ref{s3d}c at larger
energies. Finally Fig.~\ref{s3d}d shows sv5 just after its
bifurcation from s. The morphological evolution of the face-on
projections of the 3D families resembles the evolution of the
stable s orbits as the energy increases. Both sv3 and sv5 families have
complex unstable parts, starting from $E_j \approx -0.19424$ and $E_j \approx
-0.1972$ respectively. However, as in the case of fv5, this happens at
energies where $\overline{|z|}$ is large, and thus the orbits of these
families at these energies are not significant for the local disc density.

\section{Discussion}\label{blabla}
\begin{figure}
\centerline{\includegraphics[scale=0.7] {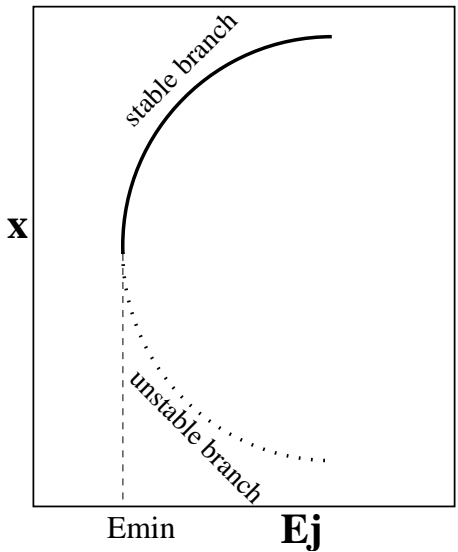}} 
\caption[]{Schematic representation of a tangent bifurcation. The
family
  starts existing for energies larger than $E_{min}$.
 }
\label{tng}
\end{figure}

In 3D Ferrers bars, families of periodic orbits beyond the gap 
at the 4:1 radial resonance, where the characteristic of the planar x1 family
has a local maximum (Fig.~\ref{f-c}),
are introduced through tangent bifurcations \citep[see][ pg.102
for a definition]{gcobook}. Tangent bifurcations are fundamental
to the study of nonlinear systems since they are one of the most
basic processes by which families of periodic orbits are created.
In 2D barred potentials they correspond to the gaps at the even
radial resonances \citep{gcog89}. In such a bifurcation a pair of
families of periodic orbits is created ``out of nothing''. They
are also referred to as saddle-node bifurcations. One of the
newborn sequences of orbits is unstable (the saddle), while the
other is stable (the node). A characteristic diagram of this type
of bifurcations is shown schematically in Fig.~\ref{tng}. The
characteristics of families f and s in Fig.~\ref{f-c} are of this
type, which means that these families are {\it not} connected to
the x1-tree. Furthermore, they build their own group of families,
i.e. their own trees.  Despite the fact that the stability is not
indicated, one can see from the shape of the characteristics that
in the 2D Ferrers bar model in \citet{ath92a} the families in the
upper part of the 4:1 resonance gap are introduced via tangent
bifurcations.

\begin{figure}
\centerline{\hspace{0cm} \includegraphics[scale=0.4]
{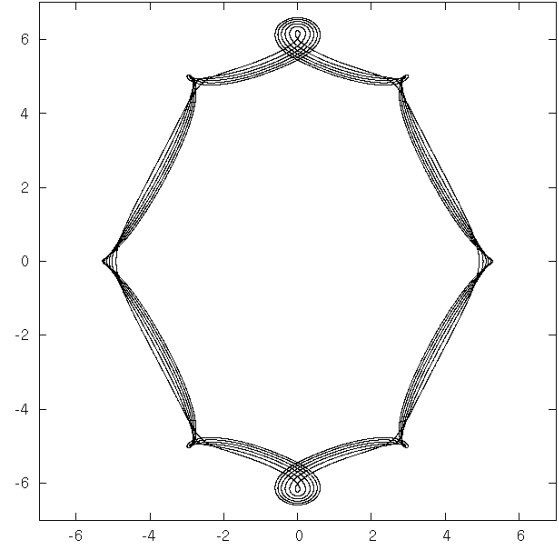}} \caption[]{A set of stable, octagonal s orbits. Note
that, while the outer boundary is
  octagonal, the inner boundary tends to a hexagonal shape.
 }
\label{octa}
\end{figure}

Close to the $E_j$ minimum at which the f and s families appear, the orbits
are characterized by 4:1 and 6:1 morphologies, respectively. As we move along
the stable branch towards higher values of $E_j$, the f orbits initially
become rounder, and then for yet larger $E_j$, they develop cusps along the
minor axis of the bar, evolving morphologically, roughly speaking, to a
hexagon.  Note that this hexagon has two sides parallel to the minor axis of
the bar. The morphological evolution of the stable s orbits forms a similar
sequence, now from a 6:1 to an 8:1 morphology, again via an ovalish shape.
Note, however, that the geometrical shape of the orbit depicted in
Fig.~\ref{s2d}a changes very fast, i.e. the s orbits have this morphology only
in a very tiny energy interval.  It evolves to a lens-type morphology and
finally to the octagonal shape we see in Fig.~\ref{s2d}c, which develops loops
along the major axis of the bar. In order to visualize a collective effect of
the presence of such octagonal orbits in the system, we plot in
Fig.~\ref{octa} successive orbits of this type selected at equally spaced
intervals in energy. This could lead to an octagonal ring or, most likely, to
a hexagonal ring with additional density enhancements close to the major axis
of the bar. The shapes of the rings that are favored due to the stability of
the planar f and s orbits can be described in general as ovals. Hexagonal
morphologies support motion parallel to the minor axis at the apocentra of the
rings.

The 2D radial bifurcations from f and s are associated with the
5:1 and 7:1 resonances, respectively. The corresponding 5:1 and
7:1 families come in pairs, with orbits symmetric with respect to
the bar major axis. Thus, although they do have stable parts, they
do not in general support a particular morphology if we consider
both of their branches. Indeed, two 5:1 symmetric periodic orbits
together represent a 10:1 morphology. In this case, non-periodic
orbits trapped around them will practically contribute to the
formation of a smooth ring around the bar. In cases where only one
of the two branches is populated, the ring will have a mainly
pentagonal or heptagonal geometry.

The existence of the 3D families is very important for the prevalence of a
ring structure around the bar, because they extend the volume of the phase
space occupied by stable ring-supporting orbits. Such families are introduced
at the S\ar U transitions of our model and some of their members are plotted
in Figs.~\ref{of3d} and \ref{s3d}. Their vertical thickness in the z direction
remains low for considerable energy intervals, at least for the orbits with
face-on projections without cusps or loops (Fig.~\ref{s3d}). This is in good
agreement with observations, since inner rings have never been seen to stick
out of the equatorial plane in galaxies which are observed edge-on. These new
families remain practically stable for energy intervals over which their
face-on projections, and the corresponding f or s `mother' families support
the ring. On the other hand, over the energy intervals where these families
are complex unstable, their face-on projections have loops and their average
height above the equatorial plane has also increased.  Since they are complex
unstable they are not associated with further bifurcating families \citep[for
definitions related with instabilities in 3D systems see e.g.][]{gcobook}. The
presence of complex instability introduces chaos in the system, however at
energies for which the families have members reaching large distances above
and below the equatorial plane. Thus their importance for the density of the
rings is already small.

The two groups of f and s families of periodic orbits include
practically all relevant morphologies of orbits one can find
between the radial 4:1 resonance and corotation. Although we can
not exclude the existence of n:1 (n$>$8) type families, we didn't
find any such orbit remaining close to the 
equatorial plane. Since, furthermore, the f and s family orbits
are sufficient on their own to account for the rings, we do not
discuss the n:1 (n$>$8) orbits any further.

\subsection{Interpretation of the ring morphologies}
\begin{figure*}
\centerline{\includegraphics[scale=0.9] {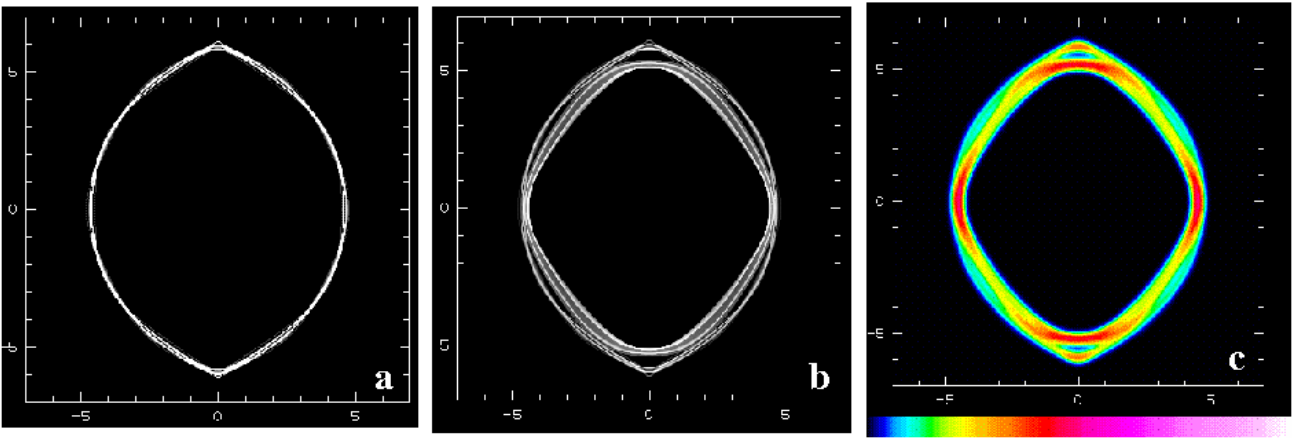}} \caption[]{
Weighted profiles obtained from stable orbits of families s, sv1
and sv3 (a) and s, sv1, sv3, f, fv1 and fv3 (b). In (c) we give a
blurred representation of (b) and we see clearly a polygonal
structure embedded in the ring. Colour in this last frame gives the projected
surface density. The orbits are chosen at the 
appropriate energy intervals in order to support an oval shaped
ring. } \label{allov}
\end{figure*}

In this section we will show that inner ring morphologies can be
built using combinations of orbits of the stable 2D and 3D
families which we described in the previous sections. Let us start
with the most common morphologies. There is a large number of
orbital combinations allowing the formation of the types of rings
illustrated in Figs.~\ref{rgal}a and b. The morphology of the s
orbits that have an outline as that shown in Fig.~\ref{s2d}b
renders well that of the most frequently encountered inner rings
(Fig.~\ref{rgal}a). This is still true if we add to them stable
sv1 and sv3 orbits, as those depicted in Figs.~\ref{s3d}a and b,
respectively. In order to demonstrate this, we give in
Fig.~\ref{allov}a the face-on profile of the weighted oval- or
lemon-shaped orbits of families s for $ -0.1990 <E_j < -0.1985$,
sv1 for $-0.1989 < E_j < -0.1968$ and sv3 for $-0.1987<E_j <
-0.1982$ and $-0.1976<E_j < -0.1958$ (sv3 is simple unstable for
$-0.1982 <E_j <-0.1976$).  These intervals have been chosen so as
to exclude all unstable orbits and all orbits with
$\overline{|z|}>0.8$. We consider orbits with $\overline{|z|}>0.8$ as
contributing little to the rings density.
We use the same technique as in
\citet*{psa02,psa03}. Namely, we first calculate the set of
periodic orbits which we intend to use in order to build the
profile. Then we pick points along each orbit at equal time steps.
The `mean density' of each orbit \citep[see][-- section
2.2]{psa02} is considered to be a first approximation of the
importance of the orbit and is used to weight it. We construct an
image (normalized over its total intensity) for each calculated
and weighted orbit, and then, by combining sets of such orbits, we
construct the weighted profile. The selected stable orbits are
equally spaced in their mean radius. The step in mean radius is
the same for all families in a figure.

In Fig.~\ref{allov}a we observe that all orbits of the s, sv1 and
sv3 families are confined inside a very narrow ring. Families sv1
and sv3 are 3D. The orbits considered here are also vertically
confined to a thin layer of a few hundreds pc having in the middle
the equatorial plane of the galaxy model. For this, and all
subsequent similar figures, we give in Tab.\ref{tab:energ} the
energy intervals from which we have taken the orbits of a family
in order to construct the profile. For building the profile we
used orbits from all the available energy intervals. Due to the
orbital crowding inside the ring, the orbits of the three families
intersect each other, and thus the corresponding gas flow will be
characterized by the presence of numerous cloud collisions in the
same area.
\begin{table*}
 \centering
 \begin{minipage}{140mm}
  \caption{Energy intervals for the stable
    orbits contributing to the weighted profiles in Fig.~\ref{allov} and
    Fig.~\ref{allex}. The successive columns give the name of the family and
    the corresponding energy intervals in the figures: \ref{allov}a,
    \ref{allov}b, \ref{allov}c, \ref{allex}a and \ref{allex}b.}
    \label{tab:energ}
  \begin{tabular}{@{}lccccc@{}}
  \hline
family & 10a & 10b & 10c & 12a & 12b \\

 \hline
 s & $-0.1990 \rightarrow -0.1985$ & $-0.1990 \rightarrow -0.1985$ &
$-0.1990 \rightarrow -0.1985$ & -- & $-0.1984
    \rightarrow -0.1972$\\
sv1 & $-0.1989 \rightarrow -0.1968$ & $-0.1989 \rightarrow
-0.1968$ & $-0.1989 \rightarrow -0.1968$ & -- &
    --\\
sv3 & $-0.1987 \rightarrow -0.1982$ & $-0.1987 \rightarrow
-0.1982$ & $-0.1987 \rightarrow -0.1982$ & -- &
    -- \\
   & $-0.1976 \rightarrow -0.1958$ & $-0.1976 \rightarrow -0.1958$ &
$-0.1976 \rightarrow -0.1958$ &  &  \\
sv5 &-- & -- & -- & -- & $-0.1984
    \rightarrow -0.1972$ \\
f &-- & $-0.2016 \rightarrow -0.1993$ & $-0.2016 \rightarrow
-0.1993$ & $-0.1990 \rightarrow -0.1978$ &
    $-0.1990 \rightarrow -0.1978$ \\
fv1 & -- & $-0.2016 \rightarrow -0.1974$ & $-0.2016 \rightarrow
-0.1974$ & -- &
    -- \\
fv3 & -- & $-0.2002 \rightarrow -0.1967$ & $-0.2002 \rightarrow
-0.1967$ & -- &
    -- \\
fv5 & -- & -- &
-- & $-0.1993 \rightarrow -0.1969$ &  $-0.1993 \rightarrow -0.1969$\\
\hline
\end{tabular}
\end{minipage}
\end{table*}

If we add, on top of this, orbits trapped around f, fv1 and fv3,
at the appropriate energy intervals, the morphology of the ring
does not change significantly. In Fig.~\ref{allov}b we add stable
orbits belonging to the families f for small energies
($-0.2016<E_j < -0.1993$), fv1 for $-0.2016<E_j < -0.1974$ and fv3
for $-0.2002 <E_j < -0.1967$. Again orbits with
$\overline{|z|}>0.8$ are excluded. The total energy width over
which the oval-building orbits exist and specify the morphology of
the inner ring is $\Delta E_j =0.0058$. The contribution of 2D
families is over a width $\Delta E_j =0.0028$, while the 3D
families support rings in an interval $\Delta E_j =0.0058$. This
indicates that 3D orbits are essential in building inner
rings. It is also evident that the width of the ring in the
face-on view is set by the contribution of orbits belonging to the
f-tree.  These orbits could be the origin of broader inner rings
observed in the near-infrared.

In Fig.~\ref{allov}c we apply a gaussian filter to the image of
Fig.~\ref{allov}b, in order to show, in a first approximation, the
shapes of secondary features which could be supported by the
orbits discussed here. The colour bar below it shows the correspondence of
colours and surface density on the ring. As we move to the right on the colour
bar 
the surface density increases. Colour helps in distinguishing the main
features supported by the orbits in our model.
It is clear that, depending on the family
which prevails, we could have the following cases: (a) If the ring
consists of s, sv1 and sv3 orbits at low energies it will be an
oval with a more or less strong lemon shape, i.e. a morphology
similar to the one of the NGC~6782 inner ring, (b) If the families
of the f-tree are dominant, they will lead to the appearance of
more hexagonal-like oval structures with corners close to the
minor axis (like in UGC~12646), (c) A combination of all
oval-supporting families, leads to the lemon-shaped oval, together
with segments parallel to the minor axis and corners close to the
minor axis of the bar. This is the morphology of the inner ring of
IC~4290 and is represented in most of its details in
Fig.~\ref{allov}c.

\begin{figure}
\centerline{\includegraphics[scale=0.5] {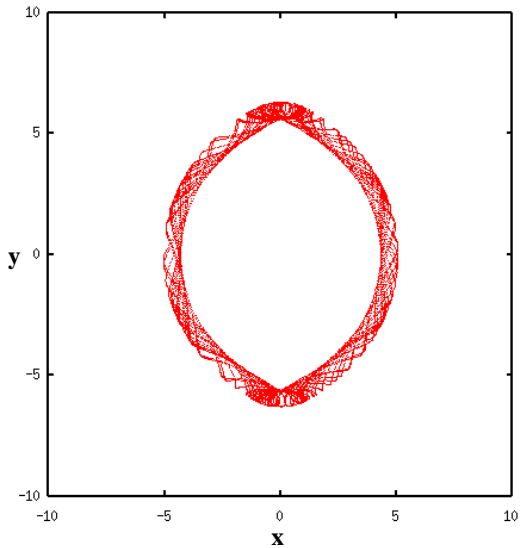}} \caption[]{
  A single non-periodic orbit trapped around the stable s periodic orbit
  at $E_j = -0.0198176$. It is integrated for more than 20
  orbital periods. It forms an oval of small width, the inner boundary of
  which is lemon-shaped.  }
\label{onelem}
\end{figure}

For the rings it is crucial that their width (in the equatorial
plane) is not too large and this is indeed the case in
Fig.~\ref{allov}c. In such a blurred image the filtering simulates
the effect of trapping around the relevant periodic orbits. To
show that the width of the gaussian that we have adopted for the
filtering is reasonable, we examined the ring structure formed by
single regular orbits, trapped around the periodic orbits
discussed above. We found that it is well confined on the face-on
view of the model, as Fig.~\ref{onelem} shows. This is due to the
nature of the characteristic diagram in the region under
consideration, since there a small deviation in the initial
conditions is sufficient to stop the orbit from being trapped
around the ring-producing periodic orbits. I.e. the trapping
around the periodic orbits, whenever it happens, is of rather
small extent. This means that the rings composed of such orbits
will have small width, in good agreement with the observations.

\begin{figure}
\centerline{\includegraphics[scale=0.45] {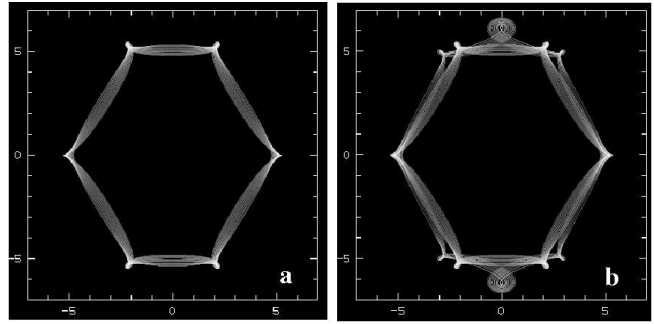}} \caption[]{
Weighted profiles obtained from stable orbits of families f and
  fv5
of the f-tree (a), and s, sv5, f, and fv5 (b). These are orbits at
larger energies than the orbits of the s, f families used for the
weighted profiles in Fig.~\ref{allov}. } \label{allex}
\end{figure}

In Fig.~\ref{allex}a we give another face-on weighted profile
consisting of stable orbits of the families f at energies $-0.1990
< E_j <-0.1978$ and fv5 at energies $-0.1993 < E_j <-0.1969$.  In
Fig.~\ref{allex}b we add also orbits from s for $-0.1984 <E_j <
-0.1972$ and sv5 for $-0.1984 <E_j < -0.1972$. This combination
gives another type of rings, which can be described as polygonal.
The orbits, as we have seen in the description of the individual
families, are of 6:1 and 8:1 types. The overall polygonal
morphology, in Fig.~\ref{allex}b, however, can be described as
hexagonal. The inner ring of NGC~3351 could be explained based on
these families.

The morphologies in Figs.~\ref{rgal}c,d can also be explained with
the stable orbits of the region, but only as specific cases,
contrarily to the many possible combinations of orbits we can use
for building the inner ring morphologies in Figs.~\ref{rgal}a,b. A
5:1 morphology (Figs.~\ref{rgal}c) can be explained by combination
of orbits belonging to fr1 (Fig.~\ref{of2d}c), or to fv1.1
(Fig.~\ref{of3d}), or to both. There are, however, two problems.
First, as we noted before, we can have a pentagonal symmetry when
only one of the two symmetric branches of the family is populated,
which presumably necessitates specific initial conditions. The
second is that these families have small stability regions. Both
are bifurcated as stable and become unstable for an energy close
to the bifurcating point. For these two reasons pentagonal rings
like in the case of NGC~3367 should be rare.

Hexagonal rings with cusps on the major axis of the bar can be
formed by material trapped around periodic orbits near the energy
minimum of the tangent bifurcation which brings family s in the
system. Nevertheless, stable s orbits with the morphology shown in
Fig.~\ref{s2d}a exist only in a very narrow energy interval. The
transition to the morphology of Fig.~\ref{s2d}b happens very fast,
while the orbit in Fig.~\ref{s2d}e is an example from the unstable
branch of the tangent bifurcation, which remains unstable for a
large energy interval. Indeed, these orbits become stable only
when they have developed large loops. The rarity of inner rings of
the hexagonal type seen in NGC~7020 \citep{bu90} reflects the way
family s is introduced in the system through the tangent
bifurcation. The stable branch of this bifurcation is associated
with the formation of ovals, while the unstable one is associated
with `NGC~7020-type' hexagons. The morphology of the stable and
the unstable orbits is very similar only very close to the energy
minimum of the s characteristic. Thus, the hexagonal geometry we
observe in NGC~7020 can be due only to non-periodic orbits trapped
in this small fraction of the phase space. Note that the inner
side of many 2D and 3D banana-like orbits trapped around the
Lagrangian points L$_4$ and L$_5$, considered in pairs, can also
form hexagons \citep[see][-- Figs.~18 and 19]{spa02a}. They can
not, however, explain rings, because of their orientation. If we would
consider at a given energy the two stable banana-like orbits
trapped around L$_4$ and L$_5$, respectively, we build a ring
elongated along the {\em minor} axis of the bar rather than around
the major axis \citep[Fig.2]{psa03}, which is contrary to
observations.

Finally we note that NGC~6782 gives also an example of diamond
morphology {\em
  in the bar}, due to orbits with that shape, which is not related to the
ring. This can be realized by inspection of the innermost isophote
drawn in Fig.~\ref{rgal}a. These isophotes may correspond to x1
diamond-shaped orbits.

\section{Conclusions}\label{conc}
We have studied the orbital structure of inner rings in static 3D Ferrers
bars. The stable orbits we find belong to families which are typical for barred
potentials with a type-2 gap in their characteristic at the radial 4:1
resonance. In 2D \citep{ath92a} and 3D \citep{spa02b} Ferrers bars this is the
usual type of gap we encounter. So we expect our conclusions to be valid for a
large area of the parameter space. This work is the first step in a study of
ring structures encountered in time-dependent, fully self-consistent $N$-body,
as well as in gaseous-response models.

The main conclusions of the present study are the following:
\begin{enumerate}
\item In 3D Ferrers bars, inner rings are due to orbits belonging to families
  in the upper part of the type-2 gap at the inner radial 4:1 resonance. They
  are grouped in two orbital trees, which have as mother-families the planar f
  and s orbits. For building the rings one cannot invoke orbits
  from the x1-tree or other families. The orbits that make the rings belong in
  their vast majority to three-dimensional families of periodic orbits.
\item The 3D bifurcating families of the two groups (fv1, fv3, fv5; sv1, sv3,
  sv5) play a crucial role in the morphology of the inner rings. They have
  large stable parts and thus they increase considerably the volume of the
  phase space
  occupied by ring-supporting orbits. The energy width over which we can find
  stable 3D orbits supporting the rings is larger than the corresponding
  interval of 2D stable families.
\item The prevailing types of inner rings are variations of oval shapes and
  are determined by the way the f and
  s families are introduced in the system, i.e. by the tangent bifurcation
  mechanism. The orbits on the stable branch of their characteristic, together
  with their stable 3D bifurcations, support ovals with a more or less strong
  lemon shape, or
  oval-polygonal rings with `corners' along the minor axis of the bar. These
  types of inner rings represent frequently observed morphologies.
\item Pentagonal rings are rare because the families building them (fr1, fv1.1)
  have small stable parts and usually come in symmetric pairs. Thus, in order
  for these rings to appear, the symmetry must be broken and only one of the
  two branches be populated due to some particular formation
  scenario. Furthermore, considerable material should be on regular
  non-periodic orbits trapped around stable periodic orbits existing only in
  narrow energy ranges.
\item If orbits are trapped around stable s periodic orbits at
  the energy minimum of the s characteristic, then an NGC~7020 morphology can
  be reproduced. Although such a morphology is in principle possible, it
  should be rare, because it would necessitate that considerable amount of
  material be on regular orbits trapped around periodic orbits in a very
  narrow energy interval. Indeed the hexagonal orbits with cusps on the {\em
    major} axis are on the unstable branch of the tangent bifurcation.
\end{enumerate}

\section*{Acknowledgments}

We acknowledge fruitful discussions and very useful comments by
G.~Contopoulos and A.~Bosma. We also thank an anonymous referee for remarks
and suggestions, which improved the paper.
This work has been supported by the
Research Committee of the Academy of Athens.  Ch.~Skokos was
partially supported by the Greek State Scholarships Foundation
(IKY).  A large fraction of the work contained in this paper was
done while P.A.P and Ch.S. were in Marseille. They thank the
Observatoire de Marseille for its hospitality. The final draft of
this paper was written while E.A. was in I.N.A.O.E. She thanks the
I.N.A.O.E. staff for their kind hospitality and ECOS-Nord/ANUIES
for a travel grant that made this trip possible.


\begin{thebibliography}{99}
\bibitem[\protect\citeauthoryear{Athanassoula}{1992a}]{ath92a} Athanassoula 
E., 1992a, MNRAS, 259, 328
\bibitem[\protect\citeauthoryear{Athanassoula}{1992b}]{ath92b} Athanassoula 
E., 1992b, MNRAS, 259, 345
\bibitem[\protect\citeauthoryear{Athanassoula et al.}{1982}]{athea82} 
Athanassoula E., Bosma A.,
  Cr\'ez\'e M., Schwarz M. P., 1982, A\&A, 107, 101
\bibitem[\protect\citeauthoryear{Buta}{1986}]{bu86} Buta R., 1986, ApJS, 61, 
609
\bibitem[\protect\citeauthoryear{Buta}{1990}]{bu90} Buta R., 1990, ApJ, 356, 87
\bibitem[\protect\citeauthoryear{Buta}{1995}]{bu95} Buta R., 1995, ApJS, 96, 39
\bibitem[\protect\citeauthoryear{Buta}{1999}]{bu99} Buta R., 1999, Ap\&SS, 
269/270, 79
\bibitem[\protect\citeauthoryear{Buta}{2002}]{bu02} Buta R., 2002,
in Athanassoula E.,  Bosma A., Mujica R., eds., ASP Conference
Series, Vol. 275, Disk of Galaxies: Kinematics  Dynamics and
Perturbations, p. 185
\bibitem[\protect\citeauthoryear{Buta \& Combes}{1996}]{buco96} Buta R., 
Combes F., 1996,
  Fund. Cos. Phys., 17, 95
\bibitem[\protect\citeauthoryear{Buta \& Purcell}{1998}]{bupu98} Buta R., 
Purcell G.B., 1998, AJ,
  115, 484
\bibitem[\protect\citeauthoryear{Buta, Purcell \& Crocker}{Buta et al.}{1996}]
{buea96}
Buta R., Purcell G.B., Crocker D. A., 1996, AJ, 111, 983
\bibitem[\protect\citeauthoryear{Buta et al.}{1998}]{buea98} Buta R., Alpert 
A.J., Cobb M.L.,
  Crocker D.A., Purcell G.B., 1998, AJ, 116, 1142
\bibitem[\protect\citeauthoryear{Buta et al.}{1999}]{buea99} Buta R., Purcell 
G.B., Cobb M.L.,
  Crocker D.A., Rautiainen P., Salo H., 1999, AJ, 117, 778
\bibitem[\protect\citeauthoryear{Buta et al.}{2001}]{buea01} Buta R., Ryder 
S.D., Madsen G.J.,
  Wesson K., Crocker D. A., Combes F., 2001,  AJ, 121, 225
\bibitem[\protect\citeauthoryear{Contopoulos}{1980}]{gcoxx} Contopoulos G. 
1980, A\&A, 81, 198
\bibitem[\protect\citeauthoryear{Contopoulos}{1983}]{gco83} Contopoulos G.
1983, Physica D, 142
\bibitem[\protect\citeauthoryear{Contopoulos}{1988}]{gco88} Contopoulos G.
1988, A\&A, 201, 44
\bibitem[\protect\citeauthoryear{Contopoulos}{2002}]{gcobook} Contopoulos G., 
2002, Order and Chaos in
  Dynamical Astronomy, Springer-Verlag Berlin Heidelberg
\bibitem[\protect\citeauthoryear{Contopoulos \&  Grosb{\o}l}{1989}]{gcog89} 
Contopoulos
G., Grosb{\o}l P., 1989, A\&A Rev., 1, 261
\bibitem[\protect\citeauthoryear{de Vaucouleurs \& Buta}{1980}]{dvb} De 
Vaucouleurs G., Buta R.,
  1980, AJ, 85, 637
\bibitem[\protect\citeauthoryear{Miyamoto \& Nagai}{1975}]{mina75} Miyamoto 
M., Nagai R., 1975, PASJ, 27, 533
\bibitem[\protect\citeauthoryear{Patsis, Skokos \& Athanassoula}{Patsis et al.}
{2002}]{psa02} Patsis P.A., Skokos
Ch., Athanassoula E. 2002, MNRAS, 337, 578
\bibitem[\protect\citeauthoryear{Patsis, Skokos \& Athanassoula}{Patsis et al.}
{2003}]{psa03} Patsis P.A., Skokos
Ch., Athanassoula E., 2003, MNRAS, 342, 69
\bibitem[\protect\citeauthoryear{Pfenniger}{1984}]{pf84} Pfenniger D., 1984, 
A\&A, 134, 373
\bibitem[\protect\citeauthoryear{Sandage \& Bedke}{1994}] {sabe94} Sandage A., 
Bedke J., 1994,
    The Carnegie Atlas of Galaxies, Vol.II., Carnegie Institute of Washington
  with the Flintridge Foundation, Washington D.C.
\bibitem[\protect\citeauthoryear{Schommer \& Sullivan}{1976}]{ss76} Schommer 
R.A., Sullivan
  T.W., 1976, Astrophys. Lett., 17, 191
\bibitem[\protect\citeauthoryear{Schwarz}{1981}]{sch81} Schwarz M.P., 1981, 
ApJ, 247, 77
\bibitem[\protect\citeauthoryear{Schwarz}{1984a}]{sch84a} Schwarz M.P., 1984a, 
MNRAS, 209, 93
\bibitem[\protect\citeauthoryear{Schwarz}{1984b}]{sch84} Schwarz M.P., 1984b, 
PASAu 5, 464
\bibitem[\protect\citeauthoryear{Schwarz}{1985}]{sch85} Schwarz M.P., 1985, 
PASAu 6, 205
\bibitem[\protect\citeauthoryear{Skokos et al.}
{2002a}]{spa02a} Skokos Ch., Patsis P.A., Athanassoula E., 2002a, MNRAS, 333, 847 
\bibitem[\protect\citeauthoryear{Skokos et al.}
{2002b}]{spa02b} Skokos Ch., Patsis P.A.,
  Athanassoula E., 2002b, MNRAS, 333, 861



\end{thebibliography}
\end{document}